\documentclass{article}
\usepackage{graphicx}
\usepackage{amsmath}

\begin{document}
 
\baselineskip 24pt

\begin{center}
{\Large {\bf  Beyond Inflation: A Cyclic Universe Scenario}}
 \vskip5mm 
Neil Turok$^{1,}$\footnote{Talk given at the Nobel Symposium `String Theory and 
Cosmology', Sigtuna, August 14-19,
2003.}
and Paul J. Steinhardt$^{2}$ \vskip3mm \mbox{}%
$^{1}$ Centre for Mathematical Sciences, Wilberforce Road, Cambridge CB3 0WA, U.K.

\mbox{}$^{2}$ Department of Physics, Princeton University, 
Princeton NJ 08544, U.S.A.

\bigskip
\begin{abstract}
Inflation has been the leading early universe scenario for
two decades, and has become 
an accepted element of the successful `cosmic concordance' 
model. However, there are many puzzling features of the 
resulting theory. It requires both high energy
and low energy inflation, with 
energy densities differing by a hundred orders
of magnitude. The questions of why the 
universe started out undergoing high energy inflation, and
why it will end up in low energy inflation, are unanswered. 
Rather than resort to anthropic arguments, we
have developed an alternative cosmology, the cyclic universe\cite{cyclic},
in which the universe 
exists in a very long-lived attractor state determined by the 
laws of physics. The model 
shares inflation's phenomenological 
successes 
{\it without} requiring an epoch of high energy inflation.
Instead, the universe is made homogeneous and flat, and 
scale-invariant adiabatic perturbations are generated
during an epoch of low energy acceleration like that seen
today, but preceding the last big bang. Unlike inflation, 
the model {\it requires} low energy acceleration 
in order for a periodic attractor state
to exist.
The key challenge facing the scenario is
that of passing through the cosmic singularity at $t=0$. 
Substantial progress has been made at the level of linearised
gravity, which is reviewed here. The challenge of extending
this to nonlinear gravity and string theory remains.
\end{abstract}
\end{center}

\section*{1. Introduction}

Observational cosmology has made tremendous progress 
over the last few years.
Many of the traditional debates have now been settled, and we 
have converged on what seems to be 
a good phenomenological description 
of the observed universe.
The universe is nearly
flat, as expected from the simplest inflationary models\cite{inflation}, 
and the fluctuations seem to be primordial, Gaussian, linear, 
adiabatic and nearly 
scale-invariant, which was 
again the expectation from inflation\cite{inflapert}. Less comforting
was the discovery that today's universe possesses
positive vacuum energy, a hundred orders of magnitude smaller than 
the vacuum energy needed to drive inflation. So any 
inflationary model not only needs the usual fine-tuning (at a level of
$10^{-10}$ or so) required to produce density variations at
the right level, but also needs fine-tuning of $10^{-100}$ or so to
get today's vacuum energy right. 

The successes of inflation have helped people to forget its
failures. 
Cosmology's greatest
conundrum,  the initial singularity - the beginning
of time and the origin of everything - remains as puzzling as
ever. Proposals for avoiding an initial singularity, 
for example the no boundary proposal, do not seem 
successfully to predict the occurrence of inflation. In
order to rescue the theory, some 
resort to anthropic arguments. However, it is questionable whether
anthropic selection is really
powerful enough to select a universe like ours.

Recently, it has been proposed that superstring theory vacua
form a vast `discretuum' within an immense landscape, 
and that inflation plus the anthropic principle
may determine our observed vacuum.  However, if we extrapolate
back in time we always encounter the big bang singularity
and the problem of how and why our region of the 
universe started out at some particular point on the landscape.
These problems 
provide strong motivation for seeking an 
alternative to inflation, which deals with the cosmic
singularity instead of assuming the universe started out
just after it.
Perhaps 
our failure to deal properly with the cosmic singularity
is the source
of the current unsatisfactory reliance upon ad hoc initial
conditions. 
A more complete dynamical
theory of the universe might yield physical 
attractor states to which the universe would be drawn
so that we did not have to start it off `by hand' in 
an arbitrary 
inflating state. 

It seems to us that the time is right to try to do better.
String theory and M theory encourage us to go beyond 
conventional field theory and
gravity and to try and make sense of singularities.
Within this framework, if consistent laws
for passing through singularities can be uncovered,
it seems entirely reasonable to expect that 
there was a universe before
the singularity.
Encouraged by the fact that the big bang singularity is 
much milder in brane world models, we have been developing
an alternative to inflation. The basic idea is that
the big bang was caused by the collision of two 3-branes 
colliding in a fourth spatial dimension. In the background
solution, the 3-volumes of the branes remain finite at the
collision even though this moment is the big bang singularity
within the conventional Einstein gravity description.

The first version of
our alternative scenario, called the Ekpyrotic universe\cite{ekpyrotic},
introduced a new mechanism for the production of 
scale-invariant adiabatic perturbations, in a pre-big bang epoch.
The second version, the Cyclic universe\cite{cyclic}, exploited the 
currently observed dark energy to solve the flatness and homogeneity
puzzles, again before the big bang. This new cosmology 
provides a surprisingly complete picture of cosmic history 
in which high energy inflation is no longer required.
Instead, the universe undergoes a periodic (and 
classically, eternal) 
sequence of big bangs and big crunches. Each bang results in
baryogenesis, dark matter formation,
nucleosynthesis and galaxy formation, and ends 
with an epoch of 
low energy acceleration of the type we see today.
The decay of the dark energy  
generates both the energy
required for the next big bang,
and the scale-invariant
density perturbations required for
structure formation in the next cycle.

Unlike inflation, the cyclic model {\it requires} the presence
of low energy acceleration, in order for the cycles to repeat.
It offers the possibility of a physical (as opposed to
anthropic) explanation for the value of today's cosmological
constant, with its incredibly small        
value in Planck units, as  
that required to sustain the attractor state. The 
cyclic attractor may also be involved in selecting the
values of other presently unexplained 
constants including the baryon/dark matter/dark energy
ratios.

The main challenge facing the cyclic scenario is 
that of tracking the state of the universe through the 
big crunch/big bang transition. As mentioned, 
the background
density and curvature of the four-dimensional universe 
remain finite all the way to the 
singularity. However, 
the big crunch/big bang transition\cite{bcbb}
is still singular because a fifth dimension (the gap between the
branes) collapses to
zero for a moment. The scenario is
built upon the conjecture that 
this dimension immediately
reappears, in a manner consistent with energy
and momentum conservation suitably defined\cite{cyclic}.

Can one define 
a consistent matching rule or S-matrix for fields and 
fluctuations across such singularities? As we discuss 
below, there is no problem in finding such
a continuation 
for free quantum fields\cite{tolley} and strings\cite{craps}. Recently,
we extended this to linearised
gravity, showing how scale invariant, adiabatic 
growing mode perturbations established before the singularity
propagate into the hot big bang\cite{newtolley}. The resulting models
yield an acceptable phenomenology for a broad
range of parameters\cite{khourynew}.
In fact, the
perturbations in the cyclic and inflationary
scenarios exhibit a surprising `duality'\cite{duality,dual2},
and the levels of tuning required to fit the observations
are almost identical in the two models\cite{Khoury}.

The  challenge of extending the continuation
across the singularity to 
nonlinear gravity remains.
At the end of this review we 
review this problem, and why we are optimistic that it can
be solved.

\section*{2. Brief Overview of the Cyclic Scenario}

\begin{figure} [!hbp]
\begin{center}
\includegraphics[clip=true,scale=0.7]{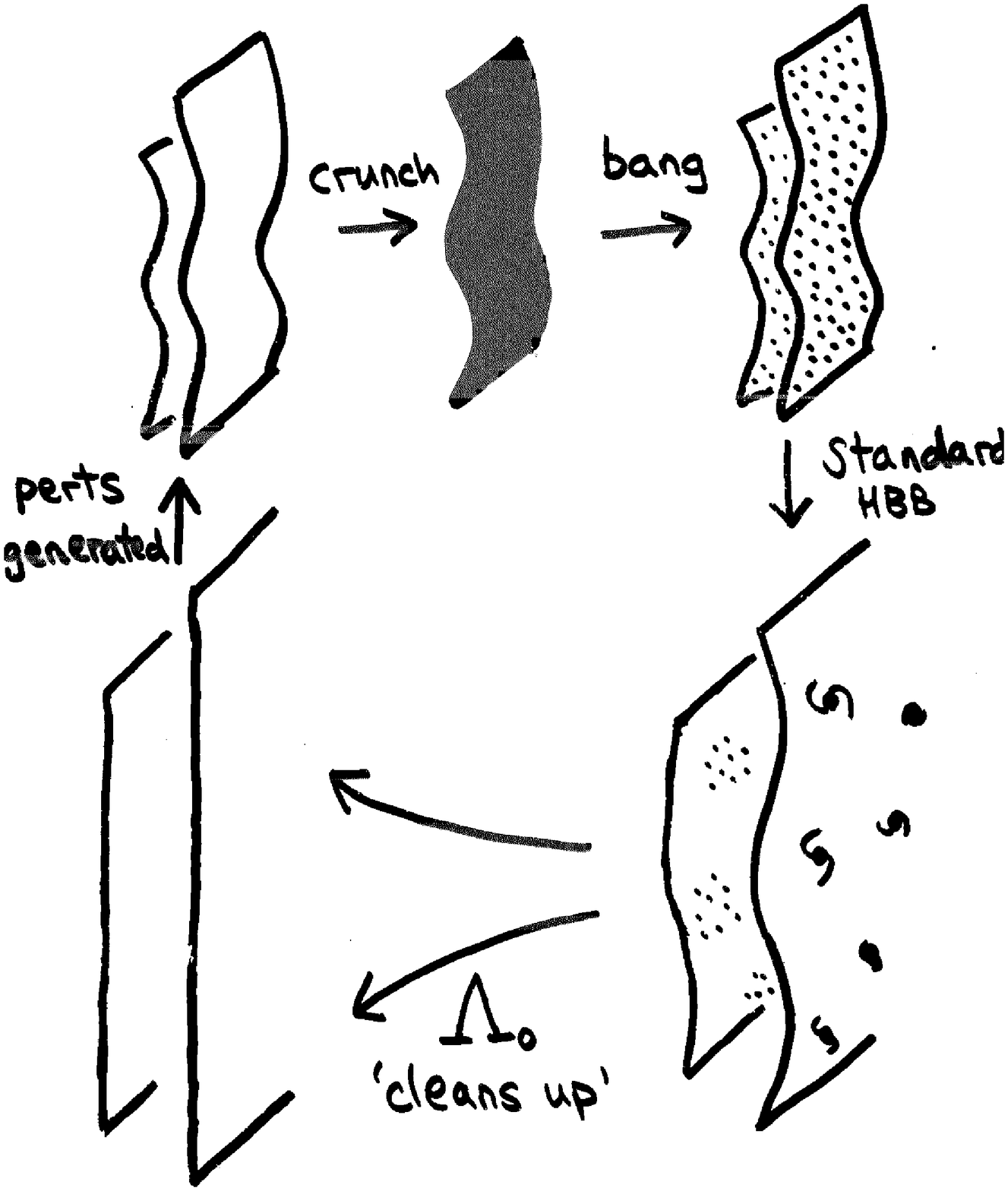}
\end{center}
\bigskip
\caption{A cartoon showing cosmic evolution in the cyclic scenario.}
\label{cyc}
\end{figure}

A cartoon representation of the scenario is shown in 
Figure \ref{cyc}. We start with 
today's universe on the lower right of the figure,
assumed to consist of an M-theory setup with 
two parallel $Z_2$ branes separated by a small
fifth dimension. Matter is confined to the branes -
perhaps with 
baryonic matter on one, and the dark matter on the other.
Positive vacuum energy 
causes the branes to expand exponentially in
the noncompact directions, `cleaning up' the universe
by diluting away perturbations, galaxies, and other nonlinear
structures
to negligible levels. The flat homogeneous
universe is a stable attractor and the 
horizon and flatness puzzles are solved much as they
are in high energy inflation except over very much larger
timescales - tens of billions of years. Following 
this low energy inflationary phase, the universe is
restored to a `blank slate' (lower left), ready for the
imprinting of perturbations which will form galaxies
in the next cycle. 

\begin{figure} [!hbp]
\begin{center}
\includegraphics[clip=true,scale=0.7]{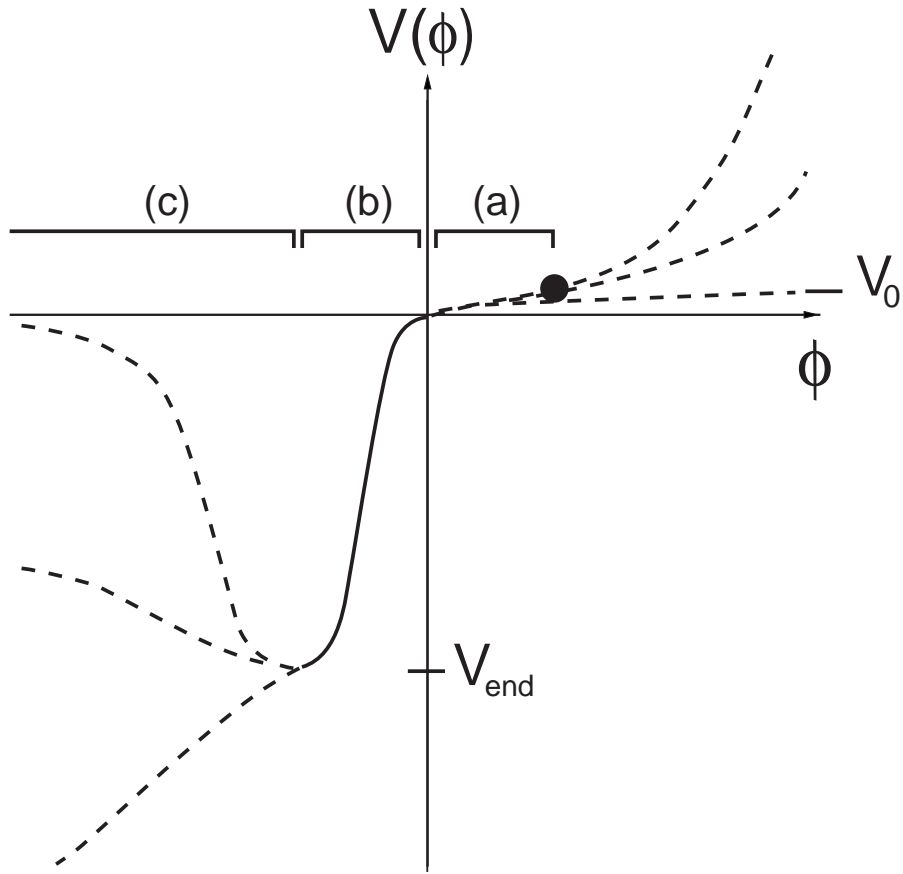}
\end{center}
\bigskip
\caption{Scalar potentials suitable for a cyclic universe model.
Running forward in cosmic time, Region (a) governs the decay of
the vacuum energy, leading to the end of the slow acceleration
epoch.
Region (b) is the region where scale invariant perturbations
are generated. In Region (c), as one approaches the big crunch
($\varphi \rightarrow -\infty$), the kinetic energy
 dominates.}
\label{pot}
\end{figure}

Now, we assume that the two branes attract one another, 
with a force which decreases rapidly as the branes 
separate (Figure \ref{pot}). Over cosmological timescales, the branes
are slowly drawn together to eventually collide. 
The long wavelength physics may be described in four-dimensional
terms, using a scalar field to represent the
inter-brane separation. The potential for this scalar
field is assumed to take the form shown in Figure 2.
In this language, the positive vacuum energy 
phase is unstable because the scalar field slowly
creeps downhill towards the region where the
scalar potential steepens and becomes large and negative. 
As we show later, in a generic class of scalar potentials
the scalar field acquires 
long wavelength, nearly 
scale-invariant perturbations as it runs downhill.

The two branes approach each other with increasing speed, 
and finally collide
at a big crunch. In the background
solution, the density and spacetime 
curvature of the branes are finite at the crunch even though
there is a singularity when
the fifth dimension disappears for an instant.  
When 
perturbations are present, the situation is more challenging.
Generic four-dimensional perturbations, {\it i.e.} the
low energy Kaluza-Klein modes, 
diverge logarithmically as one approaches
the crunch, and a prescription such as analytic
continuation is required in order to evolve them
around $t=0$ and into the post-big bang universe.
Such a continuation will be given below.
The result is that the scale-invariant perturbations
developed on the branes as they approach
propagate into the hot big bang where they
supply the adiabatic perturbations
needed for concordance phenomenology.

Note that the cyclic model has a surprising economy.
From the point-of-view of the four-dimensional
effective field theory, one 
scalar field does everything: it provides today's
dark energy, it generates the density perturbations,
regularises the singularity, and 
ignites the hot big bang. Furthermore, the five-dimensional
view of the cyclic model even explains what the scalar
field is - the size of an extra dimension - hence
it arises naturally in string and M theory models.

Is the cyclic model observationally distinguishable from inflation?
Gratifyingly, the answer is yes. High energy inflation
produces a nearly scale-invariant 
spectrum of long wavelength gravitational
waves as an inevitable side-effect. Because the
cyclic universe does not involve high energy inflation, 
no such long wavelength waves are produced. 
In the simplest inflationary models  - those with a single characteristic mass scale setting the shape of the effective potential during inflation --
gravitational waves contribute substantially
to the low multipole cosmic microwave anisotropy visible today,
at the ten or twenty 
per cent level. Furthermore, they cause a distinctive
`magnetic' signal in the polarisation of the microwave sky,
which may be measurable with
the Planck satellite to be flown by ESA in 2007. If Planck
detects magnetic polarisation, it will be a triumph for inflationary
models and will rule out the cyclic alternative. Conversely,
if Planck does not detect magnetic polarisation, it will
rule out simple inflationary models.

\section*{3. Ekpyrotic Density Perturbation Mechanism}

The ekpyrotic universe scenario\cite{ekpyrotic} introduced
a new mechanism for generating 
scale-invariant adiabatic perturbations during a contraction and
bounce.
The existence of such a simple
alternative to the usual inflationary mechanism 
was a big surprise. At heart, the 
new mechanism is 
non-gravitational and the physics behind it can be understood by 
considering a four-dimensional scalar field $\varphi$ in 
Minkowski spacetime. 

We consider a theory with a scalar potential which is
flat at large $\varphi$ but which declines steeply as 
$\varphi$ decreases.
A simple example is a
negative 
exponential,
\begin{equation}
V(\varphi)= -V_0e^{-c \varphi},
\label{eq:scalar}
\end{equation}
with $V_0$ and $c$ positive constants.
The initial conditions for the field are provided by the 
low-energy inflationary phase of the cyclic model, which
leaves $\varphi$ near its Minkowski ground state but with a 
background value 
gently rolling downwards to negative values. Gravitational
effects are subdominant since the potential and kinetic energies
are both small. The background field $\varphi_0$ satisfies 
${1\over 2} \dot{\varphi_0}^2 +V(\varphi_0) \approx
0$, which yields $\varphi\approx {2\over c} {\rm ln}(-A t)$,
with $A= c\sqrt{V_0/2}$. Here 
$t$ is negative and $t=0$ corresponds to $\varphi$ diverging.
Fluctuations in $\varphi$ may be expanded as a Fourier series 
$\delta \varphi
= \sum_{\vec{k}} \delta \varphi_{\vec{k}} e^{i \vec{k}\cdot\vec{x}}$.
The field equation reads
\begin{equation}
\ddot{\delta \varphi}_{\vec{k}} = - \vec{k}^2 \delta \varphi_{\vec{k}}
-V_{,\varphi \varphi} \delta \varphi_{\vec{k}} + O(\delta \varphi^2),
\end{equation} 
in which the mass squared term $V_{,\varphi \varphi} = c^2 V = -2/t^2$, from
the above solution. 

Those familiar with inflation will recognise the
resulting equation 
\begin{equation}
\ddot{\delta \varphi}_{\vec{k}} = - \vec{k}^2 \delta \varphi_{\vec{k}}
+{2\over t^2} \delta \varphi_{\vec{k}},
\label{eq:equation}
\end{equation}
which describes the origin of scale-invariant  perturbations
in that context (although in inflation,
$\delta \varphi$ is a conformally rescaled field
and $t$ is the conformal time). 

The initial conditions are that 
the positive frequency part of $\delta \varphi_{\vec{k}}$ is
in its Minkowski vacuum
\begin{equation}
\delta \varphi_{\vec{k}}^+ \sim {e^{-ikt}\over \sqrt{2k}}, \qquad t \rightarrow -\infty, 
\label{eq:vac}
\end{equation}
where the $(2 k)^{-{1\over 2}}$ factor is required by the usual canonical
commutation relations. For early times, each Fourier mode oscillates
under the influence of the 
$k^2$ term. However, when $|kt|$ becomes
of order unity, 
the $2 t^{-2}$ term in (\ref{eq:equation}) starts to
dominate, and the field mode evolution `freezes' into a growing
mode proportional to $t^{-1}$.
Matching the two regimes, at $|kt|\sim 1$,  
we see that
$\delta \varphi_{\vec{k}} \sim k^{-{3\over 2}} t^{-1}$ for $|kt|<<1$.
The power spectrum is the square of this amplitude and is therefore
proportional to $k^{-3}$. This is a scale invariant spectrum
with equal power in each logarithmic interval of $k$.

In fact, the time-dependence of $\delta \varphi_{\vec{k}}$
may be seen to be just that of a local 
time delay in the background solution, $\varphi(t)=
\varphi_0\left(t+\delta t(\vec{x})\right)$, or equivalently,
a scale-invariant time delay to the big crunch. 
Including gravitational back-reaction (at the linearised level) 
in the four-dimensional
effective theory has modest effect. While the perturbations
are being generated, the gravitational
fields are weak  because the energy in the scalar field
is small. And after the perturbations are generated, they
are `frozen in' as a time delay to the big crunch.
Nevertheless, there is a slight reddening of
the spectrum because the long wavelength perturbations are
generated first, and have more time to self-gravitate\cite{ekperts}.
The final result is expressible in similar terms to those
usually used for slow-roll inflation: for a general potential
of the required form, in the leading approximation including
gravitational back-reaction,
we obtain a scalar spectral index 
\begin{equation}
n_s = 1 -4 \left( (V/V_{,\varphi})^2 + (V/V_{,\varphi})_{,\varphi}\right).
\label{eq:spec}
\end{equation}
Here, as in inflation, the right hand side is to be 
evaluated as the Fourier modes concerned freeze out, 
when $|kt|$ becomes
smaller than unity. (Here and below we employ units
in which $8\pi G$ is unity).

In inflation, departures from scale invariance are
described by `slow-roll' parameters,
dimensionless measures of the first and second derivatives of 
the potential $V(\varphi$). In the cyclic model one finds 
deviations from scale invariance are similarly controlled
by `fast-roll' parameters that measure the steepness
of the potentials\cite{duality}. The 
amount of tuning required to obtain
consistency with the observations is nearly the same
in both cases\cite{Khoury}. 
A simple example of a working potential is
the form (\ref{eq:scalar}), for which $n_s-1= 4/c^2$,
so that $c>10$ is required to obtain a
scalar spectral index $n_s > 0.96$, 
compatible with the WMAP data
\cite{MAPspergel}.

\section*{Can we get through the singularity?}

In order for the scale-invariant perturbations generated 
in the pre-big bang epoch to be useful in today's
universe, we have to be able to track them across
the singularity. Whether we can do this consistently
is the key challenge to the cyclic scenario, and
most of the rest of this review will be devoted to it.

\subsection*{ Compactified Milne mod $Z_2$}

In the cyclic scenario, the singularity occurs when
the two boundary branes collide and the fifth dimension
momentarily disappears. The two branes have 
nonzero tensions (one positive and the other negative) 
and the bulk spacetime has a corresponding
warp. However, in the background solution, which possesses
cosmological symmetry,  a Birkhoff-like theorem
applies, and coordinates may be chosen in which
the bulk is static. That is, if
one sits in between the branes, there is no indication
that the branes are coming until they actually arrive. 
Therefore, in between the branes, one can choose
locally Minkowski coordinates. In fact, as $t=0$ approaches
the branes behave to leading order as 
$Z_2$ reflection planes in Minkowski spacetime (Figure \ref{Fig1}).

A model of the collision is provided
by Minkowski spacetime modded out by discrete boosts,
and then by an orbifold $Z_2$ projection.
Consider five-dimensional Minkowski spacetime with the line element
\begin{equation}
ds^2 = -dT^2 +dY^2 +d\vec{x}^2 = -dt^2 +t^2 dy^2 +d\vec{x}^2,
\label{eq:milne}
\end{equation}
where $T=t \cosh y$ and $Y= t\sinh y$. The coordinates
$(t,y)$ cover the causal past and future of the origin.
We call this half of Minkowski spacetime 
the Milne universe. We can further identify the spacetime under
discrete boosts by 
identifying $y$ with $y+L$, compactifying the $y$ coordinate and
producing the `double-cone' shown
on the right of Figure \ref{Fig1}, which we call {\it compactified Milne spacetime}. 
Finally, we impose the $Z_2$ symmetry by 
identifying 
$y\leftrightarrow L-y$ so that the branes are now the edges of the
cone shown converging for $t<0$, meeting at $t=0$ and emerging with the
same relative speed at $t>0$. We call the resulting spacetime
Compactified Milne mod $Z_2$ or ${\cal M}^C/Z_2$, and it is 
the model spacetime for
which we develop a matching rule. We will see that 
this rule the forms the basis for analysing 
more general and realistic cosmological models.

\subsection*{ Free Fields}

Classically, 
the evolution of fields on ${\cal M}^C/Z_2$ is ill-defined
for obvious reasons. There is no Cauchy surface at $t=0$, so fields
cannot be propagated across that point.
However, such
a viewpoint is certainly too naive. We are not interested in
observing the fields at $t=0$. Rather, we seek to
construct a satisfactory S-matrix relating the 
possible `in' and `out' states. There are at least three
separate ways of doing so, each of which leads to
the same result \cite{tolley}:

{\it (i) Summing boosted copies
of Minkowski plane waves}.
One can 
construct a complete basis for ${\cal M}^C/Z_2$ by summing standard Minkowski
plane waves over
boosts $y\rightarrow y +n L$, for integer $n$, and projecting
onto $Z_2$-invariant modes\cite{tolley}.
The basis so derived is defined globally 
and hence yields 
a unique propagation rule across $t=0$.

{\it (ii) Analytic continuation}. This is the most powerful
method since it extends to nonlinear fields, but it is 
also the least intuitive.
The equation of motion for the Fourier modes of a free 
fields on ${\cal M}^C/Z_2$  is
\begin{equation}
\ddot{\varphi} +{1\over  t} \dot{\varphi}  = - k^2 \varphi - {k_y^2
\over t^2} 
\varphi.
\label{phieqm}
\end{equation}
One has to continue the solutions 
around the singularity at $t=0$. The equation
is just Bessel's equation and the incoming
positive (negative) 
frequency modes of (\ref{phieqm}) are analytic in the lower
half (upper half) complex $t-$ plane. The natural continuation
of each is below or above $t=0$ respectively. 
(This is
the standard analytic continuation for Hankel functions from
positive to negative argument).

{(iii) Decomposition into left and right movers, and
real propagation in the embedding spacetime 
`around' the singularity.} 
Any solution of (\ref{phieqm}), when 
evolved towards $t=0$, splits into a sum of left 
and right moving modes. The left movers are regular on the 
left lower light cone $Y=T<0$; the right movers on the right lower
light cone $Y=-T>0$. The left/right movers therefore
have a well defined continuation into the left and
right wedges of Minkowski spacetime. A natural
assumption is that 
no other data enters these regions from past 
null infinity. Since the  
$k^2$ term in (\ref{phieqm}) behaves like a mass,
the left/right movers eventually propagate into
the future light cone, providing a unique 
propagation rule across $t=0$.

These three prescriptions turn out to yield the
same matching rule across $t=0$, which furthermore
maps the incoming adiabatic vacuum defined
as $t\rightarrow -\infty$ to the outgoing adiabatic vacuum
defined as $t\rightarrow +\infty$. Therefore
for free fields there is no particle production.
The $k_y=0$ 
modes behave as
\begin{equation}
\varphi_{\vec{k}} \sim Q+P {\rm ln} |kt|,\quad |t|\rightarrow 0,
\label{match}
\end{equation}
and the continuation rules (i), (ii) or (iii) yield
\begin{equation}
P_{out}= P_{in} \qquad Q_{out}= -Q_{in} +2 (\gamma -{\rm ln} 2) P_{in},
\label{matcha}
\end{equation}
where $P_{in}$, $P_{out}$ and $Q_{in}$, $Q_{out}$ are the constants 
appropriate to the incoming ($t<0$) and outgoing ($t>0$) field.
This result is {\it not} time reversal at the bounce,
$P_{out}= P_{in}$, $Q_{out}= Q_{in}$, which would produce the time-reversed
universe coming out. 

Picture (iii) provides helpful 
insight into the result (\ref{matcha}). 
To the left mover entering  the left wedge of Minkowski,
one must add a right moving wave, with opposite sign, in order to ensure
the field vanishes at infinity. This right mover
then enters the future light cone
leading to the sign reversal in $Q$.
This sign change is 
universal and insensitive to the warp factor and the long time
behaviour of the system far from the collision\cite{newtolley}.
The second term involving the Euler constant is 
specific to the embedding 
Minkowski spacetime and would change if we included the
effect of the warping and cosmological evolution. Happily, in the 
cosmological perturbation problem, 
$P_{in}$ is of order $k^2$ times $Q_{in}$ and hence
 negligible on large scales 
(small $k$). The sign flip of $Q$ turns out to be the key to
the propagation of large scale growing mode perturbations across
the singularity. 

\subsection*{Cosmological Perturbations}

In a recent paper\cite{newtolley}, we have analysed
the propagation of perturbations across
a big crunch/big bang collision between two boundary branes,
including general relativity at a linearised level.
The question of choosing a gauge is especially subtle in 
singular spacetimes, since 
well-defined transition functions do not exist
across the singularity. How do you then compare 
what comes in with what goes out? One requires
a common coordinate system linking two Cauchy surfaces 
in the incoming and outgoing spacetimes. 

Our procedure is to choose coordinates (or gauge)
in which each of the incoming and outgoing spacetimes 
tend asymptotically, as $t$ tends to zero, to 
perturbed
${\cal M}^C/Z_2$ in a particular gauge-fixed gauge,
in which the gravitational perturbations evolve
as massless fields on ${\cal M}^C/Z_2$. The rules
defined above 
tell us how to match such fields
across $t=0$ in the model spacetime.
For $t>0$, the perturbations on ${\cal M}^C/Z_2$ in our chosen
gauge are in one-to-one
correspondence with perturbations in the real spacetime and
we use this correspondence to map them onto the outgoing state.

In ${\cal M}^C/Z_2$, the 
modes of interest are just those of five-dimensional linearised
gravity. The $k_y=0$ modes of the five massless gravitons 
provide the four-dimensional massless fields.
Two yield four-dimensional
gravitons. Two more give a four-dimensional
gauge potential, but the $Z_2$ projection removes this.
The remaining mode is 
a four-dimensional scalar perturbation. 
This last mode is the one which is crucial for 
cosmological density perturbations. 

We need to choose a gauge in which we can treat the full,
warped spacetime, and easily relate it to the model spacetime
in which the analysis is much simpler. Much of the work
of Ref. \ref{newtolley} was to find such a gauge and
show that the matching was uniquely defined within it. In this
gauge, the four-dimensional scalar mode is represented
as
\begin{equation}
ds^2= (1+{4\over 3} k^2 \chi)(-dt^2 +t^2 dy^2)
+\left[(1-{2\over 3} k^2\chi)\delta_{ij} + 2 k_i k_j \chi\right]dx^i dx^j,
\label{ds2}
\end{equation}
and the field $\chi$ obeys the five dimensional massless
equation $\nabla^2\chi =0$, with Neumann boundary conditions
on the branes, $\partial_y\chi|_{y_\pm}=0$ (following from the $Z_2$ symmetry).
Since $\chi$ is a free field in compactified Milne mod $Z_2$,
we are able to apply the matching rule (\ref{match}) across
$t=0$. 

So much for the model spacetime. To treat the actual spacetime,
we need to include the effect of the warped bulk. We study a 
Randall-Sundrum model of type I with a positive and negative
tension brane separated by a negative $\Lambda$ bulk. We 
choose a coordinate system in which the branes are at fixed
$y=y_\pm$. The full
metric, including linearised scalar perturbations, is 
\begin{equation}
ds^2= n^2(t,y)\left(-(1+2\Phi) dt^2 - 2 W dt dy +t^2 (1-2 \Gamma) dy^2\right)
+b^2(t,y) \left( (1-2\Psi) \delta_{ij} - 2 \nabla_i\nabla_j \chi\right) dx^i dx^j.
\label{ds2a}
\end{equation}
For the background, we use the known solution 
in static coordinates (Schwarzschild-AdS). We can 
perform a coordinate
transformation to a system in which the branes are fixed ({\it i.e.} $y=y_\pm$)
as a Taylor expansion in $t$ around $t=0$. For the perturbations,
we choose a gauge where $\Gamma=\Phi-\Psi -k^2 \chi$. In this gauge,
it turns out that the $\chi$ perturbation obeys $\nabla^2 \chi =0$
(in the full five-dimensional metric),
just as it did in the model ${\cal M}^C/Z_2$ spacetime. Furthermore, there is
just enough residual gauge freedom to ensure that,
as $t$ tends to zero, the perturbed metric agrees with 
the perturbed model spacetime (\ref{ds2}) in which the matching of
free gravitational waves is
trivial. 

In the vicinity of the singularity, we solve for the background
and the perturbations as a series in $t$ and ln$|kt|$, around
$t=0$. The four-dimensional effective theory is used to 
provide the boundary data; {\it i.e.} the spacetime metrics on the
two branes. From the five dimensional Einstein equations, 
we obtain Schrodinger-like equations in $y$ at each order
in $t$, $t $ln $|kt|$, $t^2$, $t^2$ln$|kt|$, $\dots$,
which may be solved at each order by imposing the 
boundary conditions appropriate to the gauge choice made.
We have checked that we 
obtain in this way a consistent solution of five dimensional
linearised general relativity up to order $t^3$.
There is precisely enough residual gauge freedom to
choose the metric perturbations defined by 
(\ref{ds2a}) so that they take at leading order the form
(\ref{ds2}) corresponding to the model spacetime {\it i.e.}
\begin{eqnarray}
W&=&0 +O(t, {\rm ln} |kt)\cr
\Phi&=&{2\over 3} k^2 \chi+O(t, t{\rm ln} |kt)\cr
\Psi&=&{1\over 3} k^2 \chi+O(t, t{\rm ln} |kt)\cr
{\chi'\over n t}(y_\pm) &=&0 +O(t, {\rm ln} |kt),
\label{rgf}
\end{eqnarray}
as $t$ tends to zero.
This new gauge is now completely fixed, to leading order in $t$.
Therefore, by applying the rule determined for
free fields in compactified Milne mod $Z_2$, 
we obtain a unique matching rule from the 
incoming fields to the model spacetime and from the 
model spacetime to the outgoing fields. 

\subsection*{Boundary Conditions: 4d Effective Theory}

Obtaining a matching rule in the five dimensional
theory is 
only half of the story. To solve the full problem,
one has to 
follow the perturbations from their generation
well before the singularity, until they 
reach their asymptotic behaviour as $t\rightarrow 0$
in the big crunch. 
And after matching across $t=0$, one
must solve for their evolution all the way into
the late universe where we are interested
in their observational effects.

All of the work of evolving to and from $t=0$ 
is done by a very powerful technique, 
employing a four-dimensional effective theory in 
the place of the full five dimensional one. 
This changes the difficulty of the problem
from one of solving PDEs in $t$ and $y$ to 
ODEs in $t$ alone, a far easier task.
In our recent work\cite{newtolley}, we gave a new and simpler
derivation of the four-dimensional effective theory,
that shows how it reproduces the 
exact solutions for moving empty 
branes with cosmological
symmetry. When matter is present on the branes,
the four-dimensional theory is still correct 
up to $\rho^2$ terms which are consistently small 
at all times in the cyclic scenario.

The four-dimensional theory is used to 
predict the geometries of the positive and negative
tension branes, thus providing the 
boundary conditions 
for the five dimensional equations
as $|t| \rightarrow 0$. The four-dimensional effective
theory is
Einstein gravity plus a minimally-coupled scalar
field $\varphi$ representing the inter-brane separation.
The spacetime metrics $g_{\mu \nu}^\pm$ on the positive and
negative tension branes are then given by\cite{newtolley}:
\begin{equation}
g_{\mu \nu}^+ = {\rm cosh}^2(\varphi/\sqrt{6}) g_{\mu \nu}\qquad 
g_{\mu \nu}^- = {\rm sinh}^2(\varphi/\sqrt{6}) g_{\mu \nu},
\label{metrics}
\end{equation}
where $g_{\mu \nu}$ is the Einstein frame metric in the 4d theory
and we use units in which $8\pi G=1$.
Near the brane collision, the four-dimensional 
effective theory becomes singular: the Einstein
frame scale factor tends to zero as $t^{1\over 2}$ where 
$t$ is the conformal time, and the scalar field tends
to minus infinity, $\varphi \sim \sqrt{3/2} {\rm ln} |t|$,
so that both brane metrics are well behaved as $t$ tends to
zero.

In the four-dimensional effective theory, the scale-invariant
cosmological perturbations are generated by the potential
shown in Figure \ref{pot}. For $t<0$, the scale factor in the 
four-dimensional effective theory is contracting, and, 
as $t$ approaches zero, the perturbations
in the growing mode freeze out.
The four-dimensional Einstein frame metric  is, in conformal Newtonian
gauge, 
\begin{equation}
ds^2_{(4)}= a^2(t)\left(-(1+2\Phi) dt^2 +(1-2\Psi) d\vec{x}^2\right),
\label{ef}
\end{equation}
with $\Phi=\Psi$ since there are no anisotropic stresses to linear
order. The equations describing the perturbations as 
$t\rightarrow 0$ are:
\begin{eqnarray}
\delta \varphi'' +2 {\cal H} \varphi '&=& -k^2 \varphi +4 \varphi' \Phi,\cr
\Phi' +{\cal H} \Phi&=& {1\over 2} \varphi' \delta \varphi,
\label{phieqns}
\end{eqnarray}
where primes denote conformal time derivatives and ${\cal H} = a'/a$.
It is convenient to parameterise the solution in terms of the
behaviour of the comoving energy density perturbation, $\epsilon
= -{2\over 3} {\cal H}^{-2} k^2 \Phi$, which has the following
asymptotic behaviour:
\begin{equation}
\epsilon \sim \epsilon_0\left(1-{1\over 2} k^2 t^2 {\rm ln} |kt| + \dots\right)
+ \epsilon_2\left(t^2 + \dots\right),
\label{eps}
\end{equation}
and the two constants $\epsilon_0$ and $\epsilon_2$ define 
the solution for the other variables:
\begin{eqnarray}
\Phi&=& \epsilon_0\left(-{3\over 8 k^2 t^2} +{3\over 16} {\rm ln} |kt| + \dots\right) -\epsilon_2 \left( {3\over 8 k^2}+\dots\right), \cr
\varphi&=& \epsilon_0\left({3\over 8 k^2 t^2} 
+{1\over 16} {\rm ln} |kt| + {1\over 8} + \dots\right) 
-\epsilon_2 \left( {1\over 8 k^2}+\dots \right), \cr
\zeta_M&=& \epsilon_0\left(
+{1\over 4} {\rm ln} |kt| + {1\over 8} + \dots\right) 
-\epsilon_2 \left( {1\over 2 k^2}+\dots \right), 
\label{asymp}
\end{eqnarray}
where we have also given the expression for $\zeta_M$, which
is the curvature perturbation on comoving (or constant $\varphi$) 
slices. 
Note that conformal Newtonian gauge is very convenient for 
the analysis of the generation of the perturbations, and also for
relating the four- and five-dimensional theories since it is
completely gauge fixed. A further advantage is that 
one can straightforwardly read off the five dimensional
$\Phi$ and $\Psi$ perturbations on the branes from the
four-dimensional $\Phi$, $\Psi$ and $\delta \varphi$
perturbations, according to the formulas given in (\ref{metrics}).
However, the Newtonian gauge perturbations are badly divergent,
as $t^{-2}$, and do not behave as free fields in compactified Milne.
Hence it is essential to transform into a better behaved gauge 
such as (\ref{rgf})
in order to match perturbation variables across $t=0$.

Examining equations (\ref{asymp}), note that 
$\epsilon_0$ parameterises the 
the growing mode in the incoming state. In conformal Newtonian gauge,
both the
scalar field and gravitational potential possess nearly scale-invariant
perturbations {\it i.e.} $\epsilon_0/k^2$ has a nearly scale-invariant
spectrum. Conversely, 
$\epsilon_2$ is the amplitude of the decaying mode, and 
this is negligible on long wavelengths. The curvature perturbation
on $\delta \varphi=0$ slices,  
$\zeta_M$, has  
no long wavelength power.

Previous treatments of cosmological perturbations in 
singular spacetimes, for example in the pre-big bang 
scenario\cite{brus},
have for the most part simply assumed that at long wavelengths, 
$\zeta_M$ should be matched across
the singularity. In our work, we developed
an {\it ab initio} matching rule based on 
quantum field theory 
in a model spacetime which is the asymptotic limit
of the real situation. Our finding is that, with this
more sophisticated {\it and cutoff-independent} procedure, 
one does {\it not}, in general, obtain matching
of $\zeta_M$ at long wavelengths.

The reason is easy to understand. 
When we match modes
from $t<0$ to $t>0$, we should remember that,
in general, we are matching them from two unrelated coordinate
systems. It is essential to choose coordinate systems 
on either side 
so that the $t=0^-$ and $t=0^+$ time slices physically
coincide(see Figure \ref{simul}): otherwise it
makes no sense to match perturbations on them. The only common
timeslice 
is the brane collision slice: therefore one must 
choose coordinate systems for $t<0$ and for $t>0$ in which the
{\it the collision
event is simultaneous}. 
What our five-dimensional 
calculations reveal is that {\it the brane collision is
not simultaneous on $\delta \varphi =0$ slices, but it
is in our matching gauge defined above.}

The $\delta \varphi =0$ slices are natural from the
four-dimensional effective theory point of view because the
scalar field provides a natural time slicing of the 
spacetime. However, in the five dimensional
picture,  $\varphi$ only describes the brane 
separation for static branes. When the branes are
moving, there are corrections to the distance relation
depending on the brane speed
and the bulk curvature. 
Roughly speaking, the gauge choice
$\delta \varphi =0$ ensures no velocity perturbations
tangent to the branes whereas collision 
simultaneity
requires no velocity perturbations normal to the branes.
The gauge transformation required to go from $\delta \varphi=0$ gauge
to our gauge is scale-invariant in form, and introduces 
a scale-invariant curvature perturbation on the collision
surface.
Thus, in a collision-simultaneous gauge 
the spatial metrics on the branes acquire 
long wavelength, near-scale invariant
curvature perturbations at the collision. 

Finally,
let us mention some recent work confirming aspects of the above
analysis.
Craps and Ovrut \cite{craps} have analysed a string theory
$SL(2,R)/U(1)$ model which possesses a singularity 
which is locally of the same type as ${\cal M}^C/Z_2$. Through 
a group-theoretic analysis they determine a matching rule
which is fully consistent with ours in the 
limit of high momenta.
A recent analysis by Battefield et al. \cite{newbrand} including
gravitational back-reaction in the same brane model we use
has confirmed
our result for $\zeta_M$ although, unlike us, they prefer to simply 
ignore the logarithmically
divergent modes.

The result for the long wavelength curvature perturbation 
amplitude in the four-dimensional effective theory, 
propagated into the hot big bang after the brane collision 
is:
\begin{equation}
\zeta_M = { 9 \epsilon_0 \over 16 k^2 L^2} {\tanh (\theta/2) \over
\cosh^2 (\theta/2)} (\theta-\sinh \theta) \approx
- { 3 \epsilon_0 V_{coll}^4 \over 64 k^2 L^2} 
\label{emu}
\end{equation}
where $\theta$ is the rapidity corresponding to the relative
speed $V_{coll}$ of the branes at collision, and the second formula
assumes $V_{coll}$ is small. $L$ is the bulk curvature
scale, and as we saw above, $ \epsilon_0 \over 16 k^2$ has 
a scale-invariant power spectrum. The presence of radiation on
the branes before or after the collision produces an 
additional correction term given in full in
Ref. \ref{newtolley}.
As discussed above, the physical origin of the curvature
perturbation is in the time delay between the collision
timeslice and the $\delta \varphi=0$ (or comoving) hypersurfaces, on which
the branes do not possess long wavelength scale invariant
curvature perturbations.

\section*{The nonlinear regime}

We have developed and successfully 
implemented a natural prescription for
matching free fields and linear cosmological perturbations
across singularities of the type involved in the
cyclic and ekpyrotic scenarios.
However,  the
story is far from complete because we have not yet included
nonlinear effects. In the gauge we use, perturbations 
remain small until times exponentially smaller than
the Planck time, but nevertheless they generically
diverge at $t=0$ meaning that nonlinear effects
become important. We need a fully nonlinear matching
rule to deal with these. 

One useful viewpoint is provided by the standard Kaluza Klein
reduction which gives the 
four-dimensional 
Newton constant in terms of the five-dimensional Newton
constant divided by 
the size of the fifth dimension. As the latter disappears,
we are 
inexorably led into the regime of strong gravity. 

This effect is present in our linearised calculations. Even in 
the least divergent gauges, 
perturbations diverge logarithmically in time as 
$|t|\rightarrow 0$. The logarithmic divergence is 
{\it not} a four-dimensional gauge artifact, as is easily
seen\cite{ekperts}. Tensor perturbations,
which are gauge-invariant in linear theory, behave 
as free fields on a Milne spacetime and tend to 
$Q + P {\rm ln} |kt|$ as $|t|\rightarrow 0$.
For scalar
perturbations, one has only to see that 
the perturbation to the four-dimensional Ricci scalar
diverges to reach the same conclusion (for another
argument see \cite{lyth}). 

As nonlinearities become large, one should 
anticipate that perturbation theory will break down.
This does not mean that a big crunch/big bang 
transition is impossible. 
In this section
we want to explain why we believe there are grounds for
optimism. 

First, let us study the log divergence in more detail. 
Consider a $y$-independent tensor mode propagating in compactified
Milne (\ref{eq:milne}) in the
$3$ direction. Assume the diagonal mode for simplicity:
$h_{11}=-h_{22}$, with all other components zero. The equation
of motion is:
\begin{equation}
\ddot{h}_{11} + {1\over t} \dot{h}_{11} = -k_3^2 h_{11},
\label{tens}
\end{equation}
leading to the familiar behaviour 
$h_{11}= Q+P{\rm ln} |kt|$ as $|t|\rightarrow 0$. It
is not hard to see  
where this linearised behaviour leads in the 
nonlinear theory. The $k$ dependence may be absorbed
into $Q$ and $P$ and, after Fourier transforming back
to real space, one sees that the spatial dependence factorises.
Kasner solutions are
given (in any dimensions) by
\begin{equation}
ds^2 = -dt^2 + \sum_i A_i t^{2 p_i} dx_i^2,
\label{kas}
\end{equation}
where $\sum_i p_i = \sum_i p_i^2 =1$ and the $A_i$ are constants.
The background compactified Milne solution 
is $p_5=1$, $p_1=p_2=p_3=0$. A small perturbation may be 
produced by setting 
$p_1 =\epsilon <<1$ and $p_2=-\epsilon$,
and the other $p$'s are unperturbed. This solves the
Einstein equations to order $\epsilon$. To leading
order in $\epsilon$, the
exact asymptotic solution for the
metric components takes the form
$|t|^\epsilon \approx ( 1 + \epsilon
{\rm ln} |t|)$, and we recover the time dependence
of the linearised gravitational wave above.

Even though the theory is becoming nonlinear, there
is compensating simplicity near $t=0$ due to the
phenomenon of  {\it ultralocality}. As we approach the big crunch
(or the big bang), long wavelength modes cease
to be dynamical. There is just not enough time for
them to move.
To cosmologists this is familiar as
the freezeout mechanism occurring when modes leave the
Hubble horizon. In the present context, the Hubble horizon
is proportional to $t$, so all modes freeze out 
on the way to $t=0$. It is known that (in the non-chaotic
case - see below)
a generic solution of the Einstein equations tends 
asymptotically to 
the Kasner form but with $A_i$ and $p_i$ being
local functions of $\vec{x}$. From this asymptotic
solution, we can 
now see exactly how perturbation theory breaks
down. In the ultralocal limit, the exact 
solutions involve metric components behaving like $A(x) t^\epsilon(x)
= A(x) \left(1+ \epsilon(x)
{\rm ln}t +{1\over 2} (\epsilon(x)
{\rm ln}t)^2 +..\right)$. Clearly, for small enough $t$, the higher orders
dominate and the perturbative expansion in $\epsilon$ fails. 

String theory calculations\cite{seiberg}
reveal that naive string perturbation theory breaks down in
singular spacetimes like compactified Milne mod $Z_2$. 
This breakdown may be a reflection of a poor 
choice of variables
{\it i.e.}, representing the perturbed metric
as $g_{\mu \nu}^{(0)} + h_{\mu \nu}$ and expanding in $h_{\mu \nu}$.
As we have just seen, higher order terms are crucial in
representing the Kasner solution.
It seems likely to us that perturbation theory
needs to be re-summed to incorporate the effect of the
changing background as $t$ tends to zero.
A trivial example
is provided by a scalar field with an action 
$\int dt |t| \left(\dot{h}^2 -(\nabla h)^2\right) (1+h)$.
At linear order, this theory has the same $Q +P {\rm ln}|kt|$ 
behaviour we have 
been discussing. At next order, one sees that the energy density
in the field provides a diverging source term 
in the field equations:
$\ddot{h} + t^{-1} \dot{h} +k^2 h \sim \dot{h}^2$. This causes 
the breakdown of first order perturbation theory. However, a simple 
field redefinition $dH = (1+h)^{1\over 2} dh$ 
removes the interaction and 
results in a free field of exactly the type we have been
discussing. Of course, gravity is far more 
complicated than this, but the example is indicative.
The failure of
first order perturbation theory should not be interpreted to
mean that no continuation is possible in the nonlinear
theory.

Horowitz and Polchinski\cite{contrabounce} have claimed that
the breakdown of string perturbation theory signals a
disastrous instability.
Even accepting that a bounce is possible in an ideal 
background compactified 
Milne universe, they argue that the introduction of a single
particle would cause the entire space-time to collapse into a giant
black hole.  We believe this picture is misleading and that the true
story is consistent (modulo the issue of chaos, see below) 
with the locally-Kasner picture we have advocated for some time.
Let us examine this
claim within linear theory, extrapolated as
ultralocal 
Kasner evolution as $t$ approaches zero.
If we introduce into compactified Milne 
a point particle (or more accurately
a line density uniform in the fifth dimension but localised in
the noncompact directions) of mass $m$,
one can easily show using orbifold techniques that it cannot not
alter the spacetime geometry 
outside of its future light cone\cite{tts}, as one
would intuitively guess. If the particle
has existed for an arbitrarily long time, it alters the
metric at arbitrary distances, but only by a small amount.
And this metric perturbation evolves into locally Kasner
evolution as one approaches the singularity, just as
one would expect in the generic case. 
The near field of the particle yields a Newtonian potential 
$\Phi \sim - G_4 m/r
\sim - r_0^2 /(rt)$, with $r_0$ a constant, because 
the four-dimensional Newton constant is proportional
to 
$t^{-1}$.
The far field is more interesting - 
at a separation $r$ from the particle, modes freeze out
(and cease to be sourced by the particle) at a time $|t|\sim r$. 
Thereafter, these perturbations grow logarithmically
in time up to $t=0$, yielding a far field metric perturbation of the
form 
$\sim (r_0^2/r^2){\rm ln}(r/|t|)$. The ultralocal Kasner picture
tells us precisely how this extrapolates to the singularity.
The deformation of the spacetime induced by the particle
changes the original compactified Milne singularity into
a more generic Kasner singularity. It does so ultralocally, and
in a way which
has nothing to do with 
with the formation of an event
horizon or a trapped surface.

Even though the evolution of the metric becomes ultralocal,
another complication arises on the way to the singularity. 
Namely, gravitational theories can possess
a chaotic, anisotropic, Mixmaster behaviour as $t$ tends
to zero. 
At first sight, 
this would seem to be highly problematic for any putative matching
rule, since there is no simple asymptotic behaviour on either
side of the singularity. However, in recent work\cite{wesley}, we have
pointed out that in our context the problem is less severe
than one might have expected. It turns out that the potential
which we require to generate perturbations has the 
unexpected effect of suppressing anisotropies, and hence
chaos, because it produces an equation of state $w=P/\rho >>1$.
The scalar field energy blueshifts far more quickly than
anisotropic curvature and hence prevents the Mixmaster effect.
If the interbrane potential is bounded below, then eventually 
the scalar field kinetic energy comes to dominate.
In this situation, it is known that all
string theories are  borderline chaotic 
at leading order in $\alpha'$.
That is, the coupling of the 
dilaton to gauge fields is precisely the
value below which chaos ceases to exist. 

To summarise, the theory
is under excellent perturbative control all the way
to one string time before the singularity. 
First, the initial perturbations are small.
Second, the incoming state is steered
away from chaos by the $w>>1$ equation of state.
Finally, as $t\rightarrow 0$, we are within a theory which
is only marginally chaotic. One can safely conclude that there
is no chaos up to one string time 
before $t=0$. More we cannot say, until
we possess a better understanding of the
$\alpha'$ corrections. 

In the analytic continuation procedure
outlined above, it may be possible to avoid $t=0$
in the complex $t-$plane  by analytically continuing
around  the singularity from negative to positive real values of $t$
in a semicircle with radius greater than the string scale.
Then, the calculation given above
would remain essentially uncorrected by nonlinear gravitational
effects, at least on long (three-dimensional) wavelengths.
Work is currently in progress to construct such
a continuation in nonlinear gravity.

Another intriguing connection hinting at
hidden simplicity in singularity behaviour is
provided by a recent analysis of the classic 
Belinskii-Khalatnikov-Lifshitz
analysis  of cosmological singularities in the context
of string and M theory\cite{damour}.
Even though one cannot trust the effective field
theory for times smaller than the string time, 
there are 
hints that a deep mathematical structure lies
in wait, perhaps to be made use of in resolving 
the singularity. 
Studies\cite{damour} reveal the presence of 
an $E^{10}$ group thought to play a deep
role in the full spectrum of $M$ theory. Although
string theories are marginally chaotic, the 
chaos actually appears as 
a projection of geodesic motion,
suggesting an underlying simplicity which is only
beginning to be understood.

Horowitz and Maldacena have suggested\cite{hormald} an alternative
approach to dealing with future spacelike singularities.
They focus on an isolated, evaporating black hole in an
asymptotically flat spacetime, invoking  
arguments based on the AdS/CFT correspondence 
to argue that  the singularity must be reflecting. 
That is, all information in the
bulk should be stored on the flat, non-singular boundary.
So, there is no way to
lose information.  
This, by itself, is controversial\cite{preskill}, but logically consistent.
However, they further propose that a cosmological singularity may be the same,
although they are quick to note that there is no asymptotic flat, non-singular
boundary in this case.  Their argument is that the space-like cosmological
singularity is, piece by piece, like a black hole, so perhaps shares the same
property. If this interpretation is correct, it means that
perturbations can certainly not propagate through the singularity.
Time must end there, and we should impose a future boundary condition
on fields.  
We would argue that the absence of a non-singular boundary makes a
huge difference. There is no flat region where information can leak out and no
AdS/CFT-type argument that can be applied.  If information is reflected,
as they claim,
the universe could only tend to a single pre-destined final state. 
Such an extreme conclusion seems to us uncalled for. 
One would be saying that the dynamical laws of physics do not
determine the future of the universe. Instead, they require us to
prescribe
the final state. A more conservative, and therefore to us more 
plausible, alternative is that 
information, having
nowhere else to go, passes forward into the future 
through the singularity.

\section*{Causal Patch Picture and Cycling}

During the course of this meeting, we had many stimulating 
discussions with other participants regarding the relationship between 
the cyclic model and the holographic principle.   The following 
summarizes some of the discussions and our preliminary
and speculative responses.

Each cycle of our model involves
low energy cosmic acceleration, during which
the brane 
geometries expand by an exponentially large factor.
Thus the global geometry of the model is similar
to the upper half of an eternal de Sitter spacetime (Figure
\ref{dscyc}). In some respects the scenario recalls the 
steady state universe, of Fred Hoyle and co-workers. Instead
of their `C-field' which was introduced to 
continuously create matter, we have
big bangs which do so periodically. Viewed `in the large'
there are remarkable similarities between the two models.

Are the cycles eternally continuing? A naive (and perhaps
correct) argument is as follows. 
In any particular region of the 
cyclic universe, a highly improbable quantum jump 
could always occur to 
end the cycling. However,  with overwhelming probability, 
the cycling would continue in most of the universe. 
The argument is similar to that usually invoked
to justify eternal inflation. 

Stimulated by ideas of holography, a different picture of
de Sitter spacetime has been emphasised by Susskind and
others. 
In the `causal patch' approach, it is argued that only the part
of de Sitter spacetime which should be counted as 
physically relevant is the part from which an observer can
send signals to and 
receive signals from.
Furthermore, it is argued that this system should be viewed
as having a finite number of degrees of freedom, whose 
logarithm is the entropy associated with de Sitter space, 
$S_{DS} \sim (M_{Pl}^4 /\Lambda) \sim 10^{120}$ where $M_{Pl}$ is the four
dimensional Planck mass and $\Lambda$ the cosmological constant.

Consider an observer in the cyclic universe, whose world
line may be taken as the right hand edge of the diagram shown 
in Figure \ref{dscyc}. 
The physical
region as defined above is the `diamond' bounded by
the $t \rightarrow -\infty$ surface of the cycles and 
the past light cone of the observer as $t\rightarrow +\infty $
(shown by the dashed line). The observer sees an apparently
infinite number of cycles, each of which produces an entropy
of the order of the entropy we see inside the Hubble
horizon today
{\it i.e.}, $\sim 10^{90}$. The observer sees this entropy produced, and
then redshifted away. However, in the causal patch picture, it is
believed that the system is closed, so the entropy is not lost.
Rather it builds up on the horizon. Since the 
maximum amount of entropy possible in de Sitter spacetime
is $S_{DS} \sim 10^{120}$, it would appear that the maximum
number of cycles which can occur is $10^{30}$. Presumably,
beyond this number the system would be restored to thermal
equilibrium and would no longer possess an arrow of time.
We conclude that if the causal patch picture is correct,
the number of cycles is limited to $10^{30}$. The cyclic 
universe would be very long-lived, but not eternal. 
 However,
it would last enough cycles to settle into a 
strong attractor state such that we 
cannot distinguish which cycle we are presently 
experiencing.  Hence, the essence of the cycle concept 
would be preserved.

An 
interesting question that was posed at this meeting 
is why doesn't the universe produce 
near-maximal allowed entropy ($10^{120}$) in just one bounce.  In
other words, why are so many 
cyclic oscillations allowed?
 It seems to us that 
there is quite a natural answer.  Namely, the 
horizon is so small at the bounce, that only a 
limited amount of entropy 
can be generated.
Compare the entropy
generated at the collision with the maximum 
entropy possible, focussing on the comoving 
region corresponding to today's Hubble volume.
At the collision, causality forbids the formation
of black holes larger than the Hubble radius at
collision, $H_c^{-1}$, so the entropy
generated at collision is bounded by $N M_c^2$ where
$N \sim (a_0/a_c)^3$ is the number of Hubble volumes $H_c^{-3}$
contained within today's comoving Hubble volume,
and $M_c$ is the mass within
a Hubble radius, at collision. The maximum entropy
possible is however 
of order the total mass squared, where we need to
redshift the mass $M_c$ to the present time.
So $S_{max} \sim 
(N M_c (a_c/a_0) )^2$. We conclude that
\begin{equation}
{S_{coll}\over S_{max}} < {T_0\over T_c}, 
\label{entbd}
\end{equation}
where we have used $a \propto T^{-1}$ to relate the scale
factors to the temperature. For collision temperatures
between $10^5$ and $10^{19}$ GeV, we see that the collision entropy
can be {\it at most} $10^{-20}$ to $10^{-30}$ of the total 
entropy allowed. We conclude that even if the collision reheat
temperature is low, the number of cycles allowed 
before we are close to restoring thermal equilibrium is
$10^{20}$ {\it at least}.

In this meeting, Banks has described a model developed with Fischler in
which the universe begins in a state filled with black holes and expands and
cools\cite{banks}.  The gas of decaying black holes produces a scale-invariant spectrum
of perturbations over a range of scales, but the range does not extend to
the horizon radius.  They suggest a modest amount of inflation during the
expanding phase to push the spectrum to the horizon scale and beyond.
However, the cyclic model represents a possible alternative.  The long
wavelength perturbations are set during the contracting phase, and
the Banks-Fischler gas of black holes can be produced at the bounce.
In this way, the two pictures might be combined.

\section*{4. Conclusion}

In proposing a new scenario
for the early universe we have deliberately
set out to challenge standard wisdom that
a high energy inflationary phase is the {\it only} way 
to solve the conundra of the hot big bang. 

The new scenario is incomplete
at present. We do not yet have a full prescription for
nonlinear matching across $t=0$. Nevertheless,
we are encouraged by the simplicity and uniqueness 
of the matching rule in linearised gravity, and by the simplifications
wrought by ultralocality. We are led to conjecture that there exists
a consistent
analytic continuation in nonlinear gravity generalising
our linearised treatment. 

If the nonlinear matching problem is solved, and cyclic solutions 
such as we discuss are allowed, an entirely new approach to the
basic problems of cosmology is opened. The
state of the universe may be determined from 
the laws of physics in much the same way as
is the equilibrium state 
in statistical mechanics. There would be neither
a need for a special initial condition, nor
one for strong anthropic selection.

\bigskip\noindent
{\large \bf Acknowledgements} We thank  Ulf Danielsson, Ariel 
Goobar and Bengt Nilsson for organising 
this wonderful meeting, in such inspiring surroundings. 
Also Andreas Albrecht, Tom Banks, Thibault Damour, Brian Greene,
Stephen Hawking, David Kutasov, and
Lenny Susskind for  
valuable discussions. Finally, we thank Andrew Tolley and Justin Khoury 
for their collaboration. 
The work of NT was partially supported by PPARC (UK), and
that of PJS by US Department of Energy Grant DE-FG02-91ER40671 (PJS).
PJS is also Keck Distinguished Visiting Professor at
the Institute for Advanced Study with support from the Wm.~Keck
Foundation and the Monell Foundation.

\newpage

\begin{figure} [!hbp]
\begin{center}
\includegraphics[clip=true,scale=0.7]{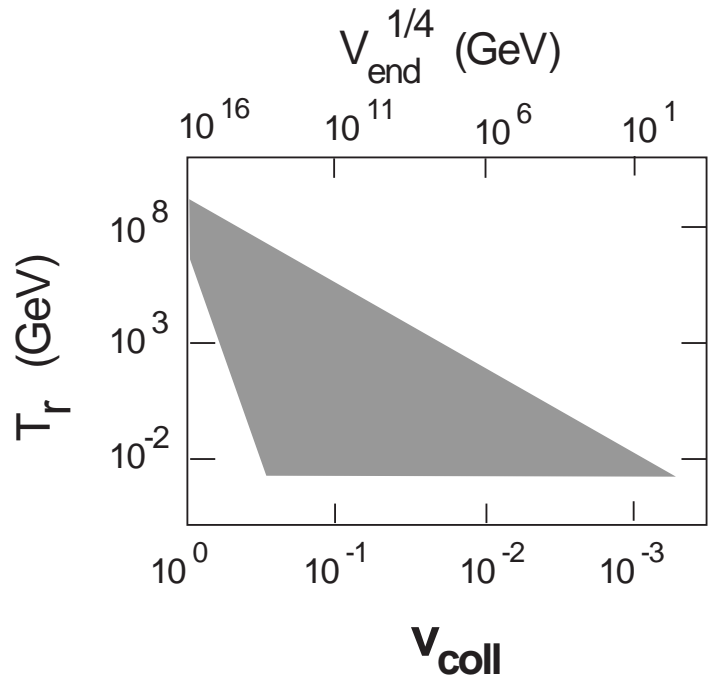}
\end{center}
\bigskip
\caption{The region of parameter space within which the
cyclic model is viable, assuming a potential of the general
form shown in Figure \ref{pot}. As well as the depth
of the potential at kinetic domination, $V_{end}$,
which determines the relative speed of the branes at
collision, $V_{coll}$, the consistency of the model 
imposes constraints on the reheat temperature at
the big bang, $T_r$. 
}
\label{limitfig2}
\end{figure}

\begin{figure} [!hbp]
\begin{center}
\includegraphics[clip=true,scale=0.5]{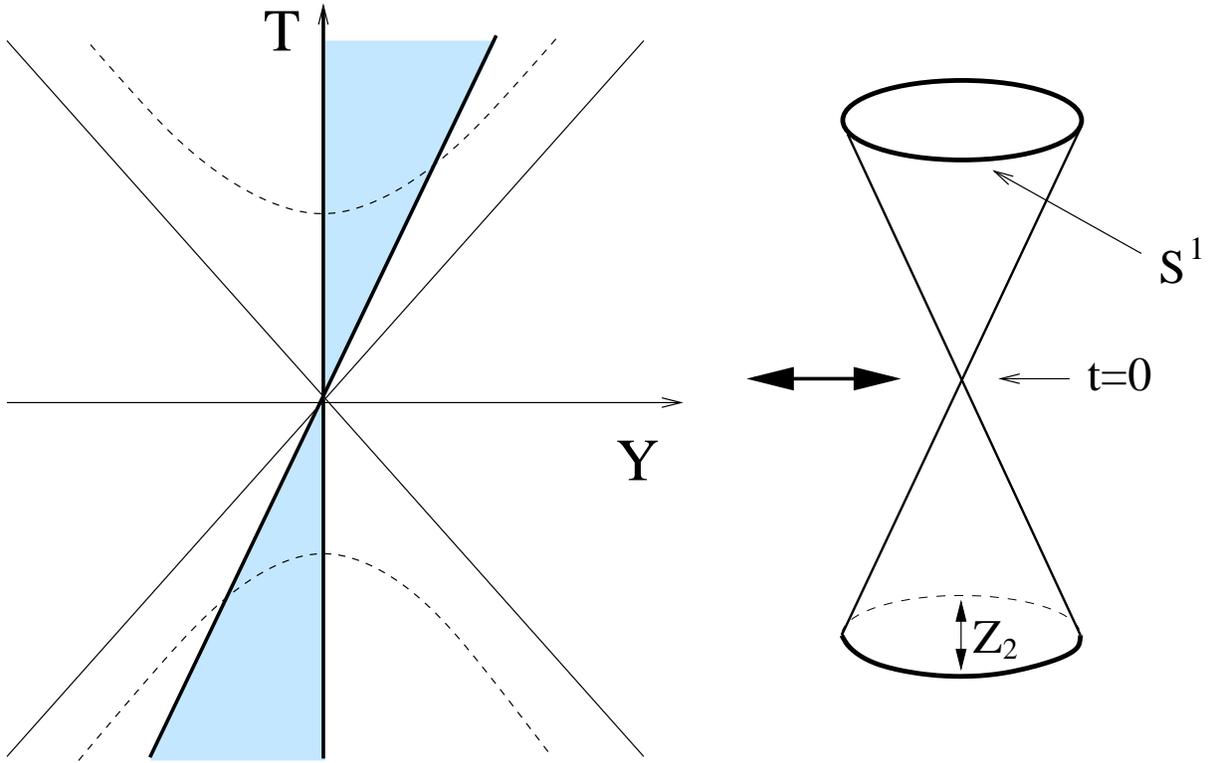}
\end{center}
\bigskip
\caption{Orbifold construction of a big crunch big bang spacetime.
One considers an embedding Minkowski spacetime, and then identifies 
the upper and lower wedges (the Milne universe) under
discrete boosts. This produces the `double-cone' at the right. 
Under a further $Z_2$ identification, one produces a spacetime
with two boundary branes approaching, colliding at $t=0$ 
and re-emerging.
}
\label{Fig1}
\end{figure}

\begin{figure} [!hbp]
\begin{center}
\includegraphics[clip=true,scale=0.5]{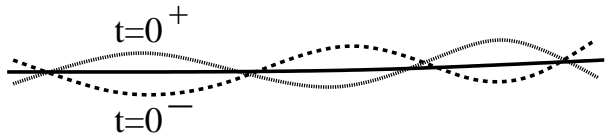}
\end{center}
\bigskip
\caption{The choice of timeslices is crucial in matching
across the big crunch/big bang singularity. The
brane collision (the $t=0$ surface, shown as a bold line) defines the natural
matching surface. It is tempting to use 
$\delta \varphi=0$ slices defined from the four-dimensional
effective theory. However, the $t=0^-$ and $t=0^+$ limits
of these
slices do not coincide with
the brane collision surface, or with each other. The incoming
$t=0^-$ surface lacks long wavelength curvature perturbations:
however, curvature is induced when one changes to the 
brane-simultaneous
collision surfaces. This curvature is transmitted to the comoving
(or constant-energy density) slices of the outgoing big bang.
}
\label{simul}
\end{figure}

\begin{figure} [!hbp]
\begin{center}
\includegraphics[clip=true,scale=0.7]{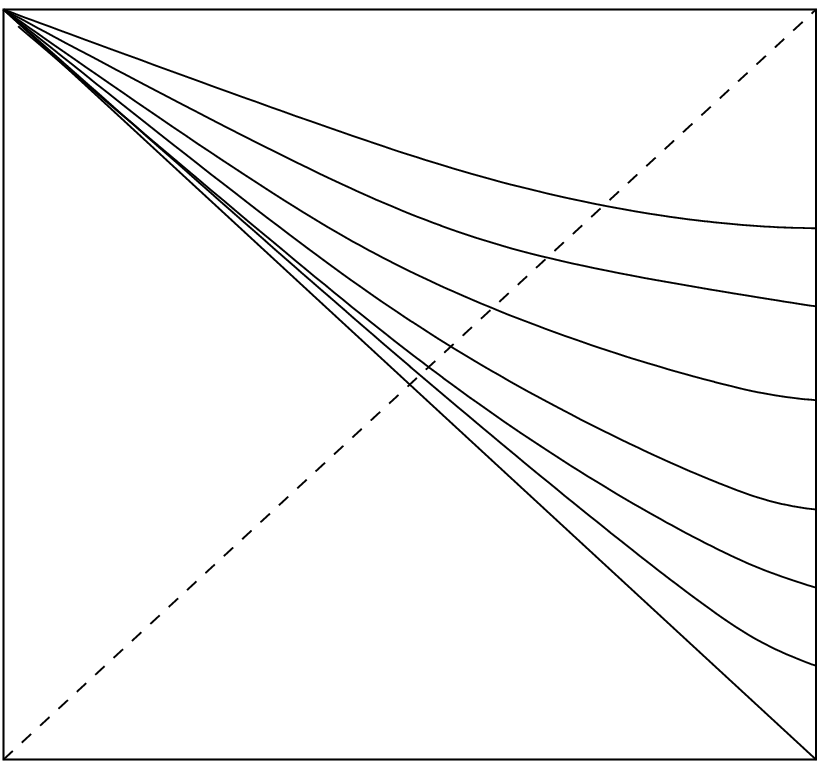}
\end{center}
\bigskip
\caption{The global causal structure of the cyclic universe spacetime.
The brane collision events define an infinite number of flat spacelike
surfaces which tend, as $t\rightarrow -\infty$, to the null
surface bisecting the diagram. As in inflation, a horizon is present,
preventing an observer from seeing the entire universe. However,
the sensitivity to initial conditions
is arguably weaker in the cyclic model than in inflation
because a very large (classically, infinite) 
number of cycles separates
us from the null initial surface. 
}
\label{dscyc}
\end{figure}


\begin{thebibliography}{99}
\bibitem{gm} The time delay is sometimes incorrectly referred to
as a gauge mode, but it is not. As long as $k \neq 0$, the time
delay mode cannot be gauged away.
\end{thebibliography}
\end{document}